\newcommand{\tends}[2]
	{\mbox{\raisebox{-8pt}{$\:\stackrel{\mbox{\large$-\!\!-\!\!\!\to$}}
	{ ^{#1 \to #2} }\:$}}}

\documentstyle[preprint,aps,epsf]{revtex}
\begin{document}

\title{Diffusion and Rheology in a Model of Glassy Materials}
\author{R. M. L.~Evans, M. E.~Cates \& P.~Sollich\footnote{Present
address: Department of Mathematics, King's College, University of London,
London, WC2R 2LS, G.B. Electronic address: peter.sollich@kcl.ac.uk}}
\address{Department of Physics \& Astronomy,
The University of Edinburgh, \\ JCMB King's Buildings, Mayfield Road,
Edinburgh EH9 3JZ, G.B. \\ Electronic address: R.M.L.Evans@ed.ac.uk,
M.E.Cates@ed.ac.uk}

\date{27 August 1998}
\maketitle

\begin{abstract}
We study self-diffusion within a simple hopping
model for glassy materials. (The model is Bouchaud's model of
glasses [J.-P. Bouchaud, {\em J. Physique I} {\bf 2} 1705 (1992)],
as extended to describe rheological properties
[P. Sollich, F. Lequeux, P. H\'{e}braud and M.E. Cates,
{\em Phys.\ Rev.\ Lett.\ }{\bf 78} 2020 (1997)].) We
investigate the breakdown, near the glass
transition, of the (generalized) Stokes-Einstein relation
between self-diffusion of a tracer particle and the
(frequency-dependent) viscosity of the system as a whole. This
stems from the presence of a broad distribution of relaxation
times of which different moments control diffusion and rheology.
We also investigate the effect of flow (oscillatory shear) on
self-diffusion and show that this causes a finite diffusivity in
the temperature regime below the glass transition (where this was previously
zero). At higher temperatures the diffusivity is enhanced by
a power law frequency dependence that also characterises
the rheological response. The relevance of these findings to
soft glassy materials (foams, emulsions etc.) as well as to
conventional glass-forming liquids is discussed.
\end{abstract}

\pacs{PACS numbers: 64.70.Pf, 66.10.Cb, 83.50.Fc}


\section{Introduction}
\label{introd}

The dynamics of systems close to a glass transition remains a
central problem in statistical physics \cite{Angell95}. Because
the glass transition is a many-body phenomenon,
models for it invariably involve some approximation and
it is important to disentangle the physical phenomena from the
approximation scheme used. In mode-coupling theories, for example,
the dynamics are dominated by a small
number of collective degrees of freedom \cite{Mode}. This is
appealing, but ignores activated processes, and obscures
the fact that glassy materials are dynamically heterogeneous with
a variety of local environments \cite{Bouchaud97}.
Another approach is to use disordered hopping models, which
treat instead single-particle degrees of freedom coupled to
a random environment, and this is the route followed here. Such
models, though unrealistic in some respects, are
likely to give a better account of those properties, such as
self-diffusion, which are not dominated by relaxation of
collective modes.

An important breakthrough in the approach based on hopping
models was that of Bouchaud \cite{Bouchaud92}, who showed that an
ensemble of {\em non-interacting} particles, moving by
thermally activated hopping at temperature $T$ in an uncorrelated
fashion through a sequence of traps, can show a glass
transition. This occurs if and only if the density of states
(the prior probability distribution of trap depths) $\rho(E)$ has an
exponential tail. Though grossly
over-simplified, this allows many properties of the model to
be calculated exactly. The exponential tail to the prior distribution
$\rho(E)$, which leads to a power-law spectrum of relaxation times and is vital
to the appearance of a glass transition, is supported by
evidence from theory on spin glasses
\cite{Sibani97} and experiments on low-temperature fluids
\cite{Rossler90b}. Such a power-law relaxation spectrum was the point of
departure for a related model of glasses \cite{Odagaki94,Odagaki95}, and
similar features appear in models of dispersive transport in disordered
semiconductors \cite{Scher91}.

The original model of Bouchaud has no explicit spatial
coordinates. Nevertheless, the extension to self-diffusion is unambiguous
(if each hop corresponds to a spatial displacement with a
well-defined second moment) and indeed Monthus and Bouchaud
\cite{Monthus96} gave an expression for the
distribution of displacements on a hypercubic lattice. For flow properties,
more choices are possible in the way spatial coordinates are
treated, but a minimal extension of the model was offered by
Sollich et al.~\cite{Sollich97,Sollich98}. This model, which we call the GR
(glassy rheology) model allows both linear and nonlinear rheological
properties to be calculated. Details of it are recalled in section \ref{GR}
below.

The model introduced in Ref.~\cite{Sollich97} was in fact proposed to
describe a class of ``soft glassy materials" (argued to
include foams, dense emulsions, etc.), in which context it was
found necessary to replace the thermodynamic temperature $T$ in
Bouchaud's model by an effective (noise) temperature $x \gg T$.
This replacement converts the GR model studied below into the
{\em soft} glassy rheology model of Refs.~\cite{Sollich97,Sollich98}, and
can be made throughout our calculations.

The glass transition in the GR model shows interesting features
which may need further explanation for readers whose background
lies in conventional glasses. Specifically, in Bouchaud's model
the glass transition $T_g$ is identified as the temperature
below which the system shows ``weak ergodicity breaking": its
Boltzmann distribution is not normalizable. This means the
system will evolve into deeper and deeper traps as time goes by,
and will never attain a steady state --- although at no finite time
are there infinite barriers partitioning phase space.  When
flow-related degrees of freedom are included in the model (see
section \ref{GR} below), one finds, as expected, that the system has a
finite elastic modulus (at zero frequency) for temperatures
$T<T_g$. Less obviously, the {\em viscosity} of the material
diverges, not at $T=T_g$ but at $T=2T_g$. Between these two
temperatures, the static modulus is zero but the viscosity infinite;
the material is what is known in rheological language as a
``power-law fluid". For applications of the model to soft glassy materials
this is a very attractive feature, since such viscoelastic behaviour is
frequently observed in these systems \cite{Sollich97,Sollich98}. In a model of
conventional glasses, on the other hand, it might be considered undesirable.
However, with the introduction of a high energy cutoff in the prior
distribution $\rho(E)$, the phenomenology of the GR model can be adapted to
model that of conventional glasses. We discuss the conceptual features of this
modification at the end of section \ref{concl}; for the actual calculations in
the present paper we restrict ourselves to the simpler case without cutoff.

Below we calculate the statistics (additional to the results of
Ref.\ \cite{Monthus96}) of self-diffusion in the GR model, and explore
quantitatively
the breakdown, near the glass transition, of the (generalized) Stokes-Einstein
relation [(G)SER] between self-diffusion of a representative particle in the
fluid and the (frequency-dependent) viscosity of the system as a whole.
(This breakdown does not, of course, imply that GSER also fails for
{\em macroscopic} probe particles which see the surrounding material as a
continuum.) We also investigate the effect of shear flow on
self-diffusion and show that this causes a finite diffusivity in the
low-temperature regime where this was previously zero.

In the next section we review the status of the GSER in both conventional
and soft glassy systems. The GR model is described in section \ref{GR} and
used in section \ref{test} to study and discuss the breakdown of the
GSER. In section \ref{analysis}, we further elucidate the physics of particle
transport in glassy systems, using both analysis and simulation of the model.
In section \ref{shear} we calculate the effects of
shear on the rate of self-diffusion, finding the diffusion constant
in the presence of continuous and periodic
shear strains of various rates and
amplitudes. In section \ref{concl} we conclude
with a discussion of our results in the context of both conventional
super-cooled/glass-forming liquids and soft glassy materials.

\section{Stokes-Einstein Relation}

The Stokes-Einstein relation (SER) between coefficients of
diffusion and viscosity (and its generalization to
frequency-dependent quantities) is of great importance to our
understanding of transport in fluids. This is especially true
since the development of new techniques to measure viscoelastic
response by light scattering from (or tracking of) probe
particles \cite{MacKintosh,Schnurr97,Mason95}. Yet the relation's range of
validity has been called into question by experiments on
super-cooled and glass-forming liquids
\cite{Fujara92,Ehlich90,Stillinger94,Cicerone96,Cicerone97,Rossler90a}.

To find the linear storage and loss moduli,
$G'(\omega)$ and $G''(\omega)$, of a
viscoelastic fluid, a rheometer is normally used to apply an oscillatory stress
to the fluid
and measure the resultant strain (or {\it vice versa}) as a function of
(angular)
frequency $\omega$. A technique has recently been developed whereby the complex
viscoelastic modulus $G^*$ ($\equiv G'+iG''$) may be
found\footnote{$G^*(\omega)$
is the Fourier transform of the time derivative of the stress relaxation modulus
$G_{r}(t)$, which relates stress on the fluid,
$\sigma(t)=\int_{0}^t dt_{1} G_{r}(t-t_{1}) \dot{\gamma}(t_{1})$, to strain rate
$\dot{\gamma}(t)$.}, even when the available sample of fluid is too small
to fill
a rheometer. The technique
involves introducing one or more neutrally buoyant probe spheres of radius $a$
into the fluid and measuring their Brownian motion. For a Newtonian fluid, which
is characterized by a single, frequency-independent viscosity $\eta$, that
viscosity is given by the familiar Stokes-Einstein formula
$\eta=k_{B}T/6\pi aD$.
Here $k_{B}$ is Boltzmann's constant and $T$ temperature. As the fluid is
Newtonian, the probe executes an unbiased random walk, and its mean-square
displacement as a function of time is $\left<r^2(t)\right>=6Dt$, which defines the
diffusion constant $D$ appearing in the Stokes-Einstein formula. In this purely
viscous fluid, $G^*$ is purely imaginary, given by
$G^*(\omega)=i\omega\eta$. As the derivative of the stress
relaxation modulus is a well-behaved function of time, its Laplace transform
$\widetilde{G}(s)$ is simply related to its Fourier transform
$G^*(\omega)=\widetilde{G}(i\omega)$. Combining the above 
expressions, we see that probe
spheres {\em in a Newtonian fluid} respect the relation
\begin{equation}
\label{GSER}
  \widetilde{G}(s) = \frac{k_{B}T}{\pi a s\left<\widetilde{r^2}(s)\right>}
\end{equation}
where $\left<\widetilde{r^2}(s)\right>$ is the Laplace transform of 
the mean-square displacement
of a probe. Equation \ref{GSER} says that the modulus of a
viscoelastic fluid is the ratio of a driving force (in this case induced by
$k_{B}T$ rather than by a rheometer) to a response (mean-square distance moved
by the sphere).

Although derived above only for the Newtonian case, Eq.\ \ref{GSER} is the
generalized Stokes-Einstein relation (GSER), and when written in this form it
applies under relatively general conditions\footnote{At sufficiently high
frequency, for a probe of finite mass, the modulus has an additional inertial
contribution which we drop.} to {\em all linearly viscoelastic fluids}, not
just Newtonian ones \cite{Schnurr97}. The basic Stokes-Einstein relation (SER)
is recovered from Eq.\ \ref{GSER} in the zero frequency limit, since viscosity
is {\em defined} by
$\eta\equiv\lim_{\omega\rightarrow 0}G''(\omega)/\omega$.
In all cases, the fluid around the probe is treated as a
featureless continuum, and therefore the result is only
valid when the sphere radius greatly exceeds the microscopic length scale
({\it e.g}.\ particle size) of the fluid. (Furthermore, probe
particles must be sufficiently dilute to ensure independent Stokesian flow
fields around each.) If, as is sometimes the case, the GSER is applied to
probe particles which are not large compared to the microscopic
scale, then it ceases to be a rigorous result. As an approximation, it is then
close to the spirit of effective medium theory (wherein a representative
particle is viewed as embedded in a continuum and its contribution to the
continuum properties then calculated self-consistently).

With macroscopic probe particles, the GSER has been fruitfully employed
\cite{MacKintosh,Schnurr97} to measure the rheological properties of
viscoelastic fluids
from light scattering or optical tracking observations of the suspended
probes. In Ref.\cite{Mason95}, Mason and Weitz applied this method
successfully to various systems, using colloidal spheres as probes.
Remarkably, in one case the medium analysed was itself a suspension of
the colloidal particles, of volume fraction $\phi=0.56$, which lies close to
the glass transition concentration. The scattering method used was
diffusion-wave spectroscopy (DWS) \cite{DWS} whose direct interpretation in
terms of self-diffusion is somewhat ambiguous; but in any case, both criteria
for the rigorous validity of the GSER were violated (no separation of length
scales, and probe particles not dilute). Yet the diffusion data for this
system, analysed as though DWS was probing self-diffusion and transformed
into putative rheological data via the GSER, compare remarkably well with
frequency-dependent rheological data measured directly. Among various
explanations that are possible for this, is the
idea that the GSER is in fact more widely valid than generally assumed
\cite{Mason95}.

On the other hand, in conventional glass-forming fluids, the (zero-frequency)
tracer diffusion and viscosity have been measured as a function of
temperature, by experiment
\cite{Fujara92,Ehlich90,Stillinger94,Cicerone96,Cicerone97,Rossler90a}
and simulation \cite{Yamamoto98}. While the SER held (within a
factor of order unity) for probes just a few times larger than the fluid
particles \cite{Cicerone96}, for equal-size probes the translational
diffusion was enhanced by orders of magnitude with respect to the SER
prediction as the glass transition was approached.

In view of the unexpected success of GSER for colloidal glassy fluids,
alongside the failure of SER for conventional
glasses\footnote{For a related discussion of the failure of Einstein
relations in biased diffusion processes, see \protect\cite{Bouchaud90}.},
a more detailed
examination of the connection between diffusion and rheology seems worthwhile,
and we pursue this in what follows.

\section{The Model}
\label{GR}

We use a simple hopping model to emulate
properties of a fluid close to its glass
transition. The model, which will be referred to as the `glassy rheology'
(GR) model, was defined in Ref.\ \cite{Sollich97}, where its response to a
macroscopic shear was calculated. In the present study we shall investigate
the relation between rheology and particle transport in the GR model.

\subsection{Definition}

The GR model is based on Bouchaud's model for glassy dynamics
\cite{Bouchaud92}, with the addition of shear strain degrees of freedom
which that model lacks. Particles in the Bouchaud model are thermally
activated at temperature $T=\beta^{-1}$ from traps of energetic
depth $E$ in a time $\tau$ which has an exponential distribution with rate
$\Gamma_{0} \exp-(\beta E)$, where $\Gamma_{0}$ is some
microscopic rate constant. Having escaped, a particle selects
its next trap {\em at random} ({\it i.e.\ }there are no spatial correlations)
from the {\em prior distribution} of trap depths $\rho(E)$. We briefly recall
the resulting behaviour of Bouchaud's model. For high temperatures $T$, the
occupancy of traps of depth $E$ evolves towards the Boltzmann distribution
$P_{\rm occ}^{\rm eq}(E)\sim \rho(E)\exp(E/T)$. As $T$ is lowered, this
distribution may cease to be normalizable, leading to a glass transition at
$T_{\rm g}^{-1}=-\lim_{E\to\infty}(\partial/\partial E) \ln\rho(E)$.
For $T<T_{\rm g}$, no equilibrium state exists, and the system shows ``weak
ergodicity breaking'' and various ageing phenomena. A finite value of
$T_{\rm g}$ implies an exponential tail in the density of states,
$\rho\sim\exp(-E/T_{\rm g})$.

The GR model ascribes some internal features to the traps.
This is done through a minimalist description in which the tensorial
aspects of the problem are ignored and the shear rate and strain are treated
as though they were scalar quantities \cite{Sollich97,Sollich98}.
The trap's potential is taken to be quadratic
in a local strain (or relative displacement) coordinate $l$, so that the
particle's energy is $\mbox{$\frac{1}{2}$} k{l}^2$,
with $k$ a constant. This applies
up to a local yield value $l_{\rm y}$, whose energy
$E=\mbox{$\frac{1}{2}$} k{l_{\rm y}}^2$ is the depth of the trap. Upon reaching this
threshold, the system locally rearranges to a new configuration, thus relaxing
the local strain $l_{\rm y}$. The
yield energy of the new potential well is drawn from the prior
distribution $\rho(E)$, and its strain $l$ set to zero. In the absence of an
externally imposed shear rate $\dot{\gamma}$, all $l$ values remain
zero, and the GR model reduces to the Bouchaud model, with hopping between
traps caused by thermal activation alone. With shear imposed, changes in
all local strains are assumed to follow the global strain rate,
$\dot{l}=\dot{\gamma}$. (This assumption is mean-field in character.) With the
particles thus dragged up their quadratic energy wells by global
shear, the barrier to thermal activation is reduced, and the local yield rate
becomes
$\Gamma_{0} \exp[-\beta(E-\mbox{$\frac{1}{2}$} kl^2)]$. If
$P_{\rm occ}(l,E;t)\,dl\,dE$
is the probability of finding an occupied trap of given
$l$ and  $E$, the above dynamics imply \cite{Sollich97,Sollich98}
\begin{equation}
  {\partial\over\partial t}P_{\rm occ} =
  -{\dot{\gamma}}{\partial\over\partial l}P_{\rm occ}
  -\Gamma_0 e^{-\beta(E-\mbox{$\frac{1}{2}$} kl^2)}\,P_{\rm occ}
  + \Gamma(t)\,\rho(E)\delta(l)
\label{EOM}
\end{equation}
The first term on the R.H.S. arises from local straining by the
macroscopic deformation. The second describes the loss of occupied traps from
the distribution by thermally activated yielding. The last term corresponds to
creation of new, unstrained configurations, whose trap depth is drawn from the
prior $\rho(E)$, at a rate equal to the total yield rate in the system,
$\Gamma(t)=
\Gamma_0\left\langle\exp[-\beta(E-\mbox{$\frac{1}{2}$} kl^2)]\right\rangle_P$.

Note that, as originally conceived in the context of soft glassy materials
\cite{Sollich97,Sollich98} each
occupied trap represented a small region of the medium, large enough for a
local shear strain $l$ to be defined, but small enough for this to be
approximately uniform within the region. However,
the model can readily be interpreted as describing the motion of individual
particles in a conventional super-cooled/glass-forming fluid, by associating
the local strain $l$
with, say, the relative displacement of a particle from the centre of a trap
formed by its neighbours.

We have not yet defined the stress in the GR model.
Due to stochastic yielding events, the local stress, which we take to be $kl$,
is inhomogeneous (it follows the local strain $l$, rather than the
macroscopic strain). Since we have stated that the strain rate $\dot{l}$ is
everywhere equal to the global strain rate $\dot{\gamma}$, these local
stresses must combine according to a straightforward average so that
$\sigma = k\left<l\right>\equiv k \int\!\, l\,P_{\rm occ}( l,E;t)\,d l\, dE$.
As discussed in section \ref{breakdown}, this corresponds to a fully
parallel mechanical circuit; this is the only combination consistent with
affine shear, and may well be approached in many physical systems.

\subsection{Rheology and Diffusion}

The dynamics described (Eq.\ \ref{EOM}) are further motivated in Refs.\
\cite{Sollich97,Sollich98}, where the GR model's constitutive relation between stress
and strain rate is calculated. We now briefly summarize the linear rheology
that the model predicts. We use non-dimensional units for time and energy by
setting
$\Gamma_0=T_{\rm g}=1$; we also re-scale our strain variables ($l,\gamma$)
and stress $\sigma$ so that $ k=1$.  The complex dynamic shear modulus
$G^*(\omega)=G'+iG''$ describes the stress response to small shear
strain perturbations around the equilibrium state. As such, it is well
defined ({\it i.e.\ }time-independent) only above the glass transition,
$T>1$.  Expanding Eq.\ (\ref{EOM}) to first order in the amplitude
$\gamma_{0}$ of an oscillatory strain $\gamma(t)=\gamma_{0}\cos\omega t$,
we find
\begin{equation}
\label{GRresult}
 G^*(\omega)=\left<\frac{i\omega\overline{\tau}(E)}
                {i\omega\overline{\tau}(E)+1}\right>_{\rm occ}
\end{equation}
where the average is taken with respect to occupied traps, and
$\overline{\tau}(E)=\exp(\beta E)$ is the mean (Arrhenius) residence time
for a trap of depth $E$. (We distinguish this from the actual
residence time $\tau$ in such a trap, which is a random variable.) Equation
\ref{GRresult} corresponds to a distribution of Maxwell modes whose spectrum
of relaxation times is given by the equilibrium distribution
$P_{\rm occ}^{\rm eq}(E)\sim\exp(\beta E)\rho(E)$. Given that $\rho(E)$
has an exponential tail (required for $T_{g}$ to be finite), the
distribution of residence times thus exhibits power-law behaviour for large
$\tau$: $\Psi_{\rm occ}(\tau)\sim \tau^{-T}$.
This leads to power laws\footnote{Such phenomenology is not confirmed for
conventional glass formers. However, with a high-energy cutoff in $\rho(E)$,
which may be apparent in conventional glasses, Maxwellian behaviour
re-appears at low $\omega$. See section \protect\ref{concl}.}
for $G^*$ in the low frequency range:
\begin{equation}
\label{powerlaws}
\begin{array}{lcllcll}
G'  & \sim &\omega^2            &\mbox{for $3<T$}, \quad
    & \sim &\omega^{T-1}        &\mbox{for $1<T<3$} \\
G'' & \sim &\omega              &\mbox{for $2<T$}, \quad
    & \sim &\omega^{T-1}        &\mbox{for $1<T<2$}
\end{array}
\end{equation}
For $T>3$ the system is Maxwell-like at low frequencies, whereas
for $2<T<3$ there is an anomalous power law in the elastic modulus.
Most interesting is the regime $1<T<2$, where $G'$ and $G''$ have constant
ratio; both vary as $\omega^{T-1}$. Moreover, the frequency exponent
approaches zero as $T\to 1$, resulting in essentially constant values of
$G''$ and $G'$. Note, however, that the ratio $G''/G'\sim T-1$ becomes small
as the glass transition is approached.

This increasing dominance of the
elastic response $G'$ prefigures the onset of a yield stress for
$T<1$~\cite{Sollich97,Sollich98}. However, for $T<1$ the linear viscoelastic moduli show
slow time evolution and ageing effects \cite{Fielding} in accordance with the
weak ergodicity breaking discussed in section \ref{introd}.
Accordingly, in the following section, where the GSER for this model is
discussed, we consider only
$T>T_g =1$. However, that does include the regime $T_g \le T \le 2T_g$
which would lie below the ``glass transition temperature" were this to be
defined operationally by the divergence of the viscosity.
As $T_{g}$ is approached from above, the equilibration time following a
quick quench diverges. Hence, even for $T>T_{g}$, non-equilibrium
situations are also of interest in glass-forming materials, though we
shall consider only equilibrium situations in the following section.

In order to study self-diffusion \cite{Monthus96}, we must clarify further the
spatial interpretation of our GR model. We shall associate each relaxation
event with a step on an unbiased random walk. We draw each step from a
Gaussian\footnote{Given that the walk is Markovian, our choice of geometry
of the individual steps will not influence the large-scale behaviour.}
of unit variance in each spatial direction (thus defining a
length unit for the model). This gives a meaning to
the square displacement, $r^2(t)$.  After each step, the randomly walking
particle is temporarily trapped within a potential well. Within each well, we
assume rapid `thermalization' of the particle, which therefore samples a
Gaussian distribution of displacements at each site on the walk\footnote{We
neglect modification of the Gaussian by the leaky boundary condition at the
edge of the well. This is consistent with the unmodified hopping probability
used to construct the GR model \protect\cite{Sollich97,Sollich98}.}.
In fact, we assume this thermalization to be
instantaneous, thus discarding information about high frequency intra-well
dynamics. Hence, we model the time regime of $\alpha$-relaxation, but not of
$\beta$-relaxation.

\section{Test of the GSER for Small Probes}
\label{test}

  We have the complex linear viscoelastic modulus for the GR model,
quoted in the section above. It is now our task to compare this with the
prediction of the GSER, applied to a probe particle which is representative of
the fluid itself. The result of Eq.\ \ref{GSER}, whose validity is in doubt for
a small probe of this kind, will be denoted $\widetilde{G}_{\rm sp}$. Calculating
$\widetilde{G}_{\rm sp}$ from Eq.\ \ref{GSER} requires a knowledge of
$\left<\widetilde{r^2}(s)\right>$
in the equilibrium fluid. The brackets $\left<...\right>$
denote an average over all random walks of the particles, in
terms of choice of path and of trap depths drawn from the prior distribution
$\rho(E)$. In the absence of intra-well structure, the treatment of
continuous-time random walks  is discussed in Refs.\
\cite{Bouchaud90,Kehr78,Haus87,Monthus96}.
The mean-square displacement can be found from the correlation
function $C$, defined by
$C(\mbox{\boldmath$q$},t)\equiv\left<e^{i\mbox{\boldmath$q$}
	\cdot\mbox{\boldmath$r$}(t)}\right>$.
This is calculable \cite{Monthus96} via
$C(\mbox{\boldmath$q$},t)=\sum_{N=0}^{\infty}\widehat{Q}_{N}(\mbox{\boldmath$q$})
	\, p_{t}(N)$
where
$p_{t}(N)$ is the probability of having performed exactly $N$ hops at time
$t$, and $\widehat{Q}_{N}(\mbox{\boldmath$q$})$ is the
Fourier transform of the positional
probability distribution after an $N$-step random walk.
As described in section \ref{GR}, the positional distribution is the
convolution of a standard random walk and a ``Gaussian of thermalization"
(whose variance is $Td/k$, where $k=1$ is the local spring
constant, and $d$ the dimension of space). So its Fourier transform
$\widehat{Q}_{N}(\mbox{\boldmath$q$})$ is the {\em product} of the usual function
$\exp(-Nq^2/2)$ and another Gaussian.

The quantity $p_{t}(N)$ depends only on the distribution of residence times.
Note that the time between hops along a random walk is drawn from the
distribution $\Psi_{\rm hop}(\tau)$, which is {\em not} the distribution of
residence times of occupied states in the system, $\Psi_{\rm occ}(\tau)$. The
former distribution gives the inter-hop times {\em available} to a random
walker upon selecting its next trap. The latter gives the residence times of
all the states in the system which are {\em presently occupied}. Thus
$\Psi_{\rm occ}(\tau)$ is related to $\Psi_{\rm hop}(\tau)$ by a weight
factor $\tau$, since the likelihood of finding a given trap occupied is
proportional to the time for which it is occupied:
\begin{equation}
\Psi_{\rm occ}(\tau) = \frac{\tau\Psi_{\rm hop}(\tau)}{\left<\tau\right>_{\rm hop}}.
\end{equation}
Accordingly, averages of some
stochastic quantity ${\cal Q}$ with respect to the two distributions are
related by
\begin{equation}
\label{averages}
  \left<{\cal Q}\right>_{\rm hop}=\frac{\left<{\cal Q}/\tau\right>_{\rm occ}}
    {\left<\tau^{-1}\right>_{\rm occ}},
  \hspace{5mm}
  \left<{\cal Q}\right>_{\rm occ}=\frac{\left<\tau {\cal Q}\right>_{\rm hop}}
    {\left<\tau\right>_{\rm hop}}.
\end{equation}
In terms of a typical random walk, $\left<{\cal Q}\right>_{\rm hop}$ is the average
with respect to the steps, while $\left<{\cal Q}\right>_{\rm occ}$ is the average
with respect to time. Note that, in calculating the probability  $p_{t}(N)$
of having performed $N$ hops at time $t$, the
time spent in each trap is drawn from $\Psi_{\rm
hop}(\tau)$ {\em except for the first trap} \cite{Kehr78,Haus87},
which is selected from $\Psi_{\rm occ}(\tau)$.
This reflects the fact that the walker is not
introduced to the system at time zero, but is
selected at random from the traps already
occupied. This careful choice of the first trap
is important to the diffusive behaviour at
early times \cite{Kehr78} and, as we shall see, its influence
persists for increasingly long times as the glass
transition is approached.

  From the definition of $C(\mbox{\boldmath$q$},t)$ it follows that, in terms of its
temporal Laplace transform $\widetilde{C}(\mbox{\boldmath$q$},s)$,
\begin{equation}
  \left<\widetilde{r^2}(s)\right> = -\left.\nabla_{\mbox{\boldmath$q$}}^2
	\widetilde{C}(\mbox{\boldmath$q$},s)\right|_{\mbox{\boldmath$q$}
	=\mbox{\boldmath$0$}}
\end{equation}
where $\nabla_{\mbox{\boldmath$q$}}$ denotes a derivative in
$\mbox{\boldmath$q$}$-space. From this prescription, we find
\begin{equation}
\label{r2s}
  \left<\widetilde{r^2}(s)\right> =
  \frac{\widetilde{\Psi}_{\rm occ}(s)\,d}{s\,(1-\widetilde{\Psi}_{\rm hop}(s))}
  + \frac{T\,d}{s}.
\end{equation}
Here, $\widetilde{\Psi}(s)$ is the Laplace transform of $\Psi(\tau)$. The
distribution of available escape times $\Psi_{\rm hop}(\tau)$ is given by
averaging the exponential distribution for a single trap over the prior
energy distribution $\rho(E)$ thus:
\begin{equation}
\label{psioftau}
  \Psi_{\rm hop}(\tau) = \int_{0}^{\infty}dE \, \rho(E) \,
  \exp-[\beta E + \tau e^{-\beta E}].
\end{equation}
So its Laplace transform is
\begin{equation}
\label{laplace}
  \widetilde{\Psi}_{\rm hop}(s) = 
	\left<\frac{1}{1+s\overline{\tau}(E)}\right>_{\rm hop}
\end{equation}
with $\overline{\tau}(E)=e^{\beta E}$ the mean residence time for trap depth
$E$. The expression for $\widetilde{\Psi}_{\rm occ}(s)$
is the same, with the average
over hops replaced by an average over occupied traps. So, from Eq.\
\ref{averages},
\begin{equation}
  \widetilde{\Psi}_{\rm occ}(s) = \frac{1}{\left<\tau\right>_{\rm hop}}
  \int_{0}^{\infty}\!\!dE\,
  \frac{\rho(E)\overline{\tau}(E)}{1+s\overline{\tau}(E)}
  = \frac{1-\widetilde{\Psi}_{\rm hop}(s)}{s\left<\tau\right>_{\rm hop}}
\end{equation}
Substituting this expression into Eq.\ \ref{r2s} cancels the $\Psi$-dependence.
So we see that the second moment of the distribution of displacements,
\begin{equation}
\label{r2s2}
  \left<\widetilde{r^2}(s)\right> = \frac{Td}{s} +
	\frac{d}{s^2\left<\tau\right>_{\rm hop}},
\end{equation}
is independent of any properties of
the distribution of trapping times except for its mean.

If we use this expression for the self-diffusion of a small probe to
calculate the modulus in Eq.\ \ref{GSER} (and substitute $s=i\omega$ to find
the complex viscoelastic modulus in the Fourier domain), we find\footnote{We
have set $d=3$ and particle radius $a=k_{B}/3\pi$ to simplify the constant of
proportionality in the GSER (Eq.\ \protect\ref{GSER}).}
that the GSER, applied to a small probe, predicts a complex modulus
\begin{eqnarray}
\label{wrong}
  G^*_{\rm sp}(\omega) &=& \frac{i\omega\tau_{0}}{1+i\omega\tau_{0}} \\
  \mbox{with }\;\;\tau_{0} &\equiv& T\left<\tau\right>_{\rm
hop}=T/\left<\tau^{-1}\right>_{\rm occ}.
                                   \nonumber
\end{eqnarray}
This is {\em exactly} Maxwellian behaviour at all temperatures for which an
equilibrium distribution exists ({\it i.e.\ }for all \mbox{$T>T_{g}=1$}).
Clearly, this expression is completely at odds with the actual rheology of the
GR model as described in section \ref{GR}.

\subsection{Breakdown of the GSER}
\label{breakdown}

 The generalized Stokes-Einstein relation,
when applied to a small probe, has failed in two distinct ways which we
now discuss. The first is a (fairly trivial) discrepancy in the temperature
dependence, which is inherent in any hopping model. To study the spurious
$T$-dependence of the characteristic time in Eq.\ \ref{wrong}, let us take
the inverse Laplace transform of Eq.\ \ref{r2s2}. Thus we find the nature of
the probe's diffusive motion as a function of time:
\begin{equation}
\label{r2t}
  \left<r^2(t)\right> = Td + td/\left<\tau\right>_{\rm hop}.
\end{equation}
The time-independent term $Td$ is the variance of the local `Gaussian of
thermalization' in which the probe finds itself at each step of its walk, and
respects equipartition of energy amongst the elastic degrees of freedom. The
second term says that diffusion is exactly linear on all time
scales\footnote{We note that the diffusive term in Eq.\ \ref{r2t} is
equivalent to the infinite-dimensional limit of the exact result for a random
trap model on a hyper-cubic lattice obtained by Schroeder
\protect\cite{Schroeder75} and Kehr et al.\ \protect\cite{Kehr78,Haus87}.},
with a diffusion constant
\begin{equation}
\label{D}
  D = \mbox{$\frac{1}{2}$}\left<\tau\right>^{-1}_{\rm hop} =
	 \mbox{$\frac{1}{2}$}\left<\tau^{-1}\right>_{\rm occ}.
\end{equation}
The GSER implies that the mean-square speed (also given by equipartition,
$\left<\frac{1}{2}mv^2\right>=\frac{1}{2}k_{B}Td$)
sets the rate of translation along
the random walk. Indeed this is why the diffusion constant is proportional
to temperature in dilute gases, where particles are not caged, so their
directions rapidly become decorrelated. Like any activated hopping process
\cite{Goldstein69} with a
temperature-independent microscopic attempt rate ({\em e.g.\ }the
Eyring model of dense fluids \cite{Eyring}), the GR model lacks such a factor
$T$ in the diffusion constant (Eq.\ \ref{D}). This is the source of the
rogue $T$-dependence of $G^*_{\rm sp}$. As the GR model is designed to be
applied close to the glass transition, factors of $T$ are of order unity.
This source of deviation from the GSER is therefore not of great importance
to most experiments, which usually only test it up to factors
of order unity \cite{Cicerone96,MacKintosh,Mason95}.

However, the GSER is violated by much larger factors than this arising from
the second source of error in Eq.\ \ref{wrong}. This is the ``misplacement" of
the thermal averaging brackets, when compared with the correct expression in
Eq.\ \ref{GRresult}. In a Maxwellian fluid, the distribution of lifetimes
is narrow, since there is a single characteristic relaxation time.
Thermal averaging then commutes with the other algebraic operations, so the
incorrect placement of the thermal brackets becomes irrelevant in that
case\footnote{In fact, the GSER for small probes is tantamount to mean field
theory, in which averaging indeed commutes with certain algebraic
operations.}. However, the GSER result breaks down for broad distributions
such as exist both in many real glass formers, and in the GR model. To
illustrate this, let us compare the diffusion constant (Eq.\ \ref{D}) with
that predicted by the Stokes-Einstein relation (zero-frequency limit of the
GSER), given the GR model's actual viscosity. From Eq.\ \ref{GRresult} with
$\eta\equiv\lim_{\omega\rightarrow 0}G''(\omega)/\omega$, the SER gives the
diffusion constant as $T/2\eta=T/2\left<\tau\right>_{\rm occ}$. The first moment of
the distribution of times for which traps are occupied $\left<\tau\right>_{\rm occ}$
diverges in the GR model for temperatures below $T=2$, whereas
$\left<\tau^{-1}\right>_{\rm occ}$, which appears in Eqs.\ \ref{D} and \ref{wrong},
remains finite. Thus $D$ is {\em enhanced} with respect to the SER value
as the viscous divergence is approached, as is widely observed in experiment
and simulation \cite{Fujara92,Ehlich90,Stillinger94,Cicerone96,Yamamoto98}.
We note, however, that the GR model is not sufficiently elaborate to account
for the power law ($1/D\propto\eta^{\xi}, \xi<1$) which has been observed to
replace the SER in some supercooled fluids \cite{Fujara92}.

The rheology of the GR model is equivalent to a set of over-damped harmonic
oscillators ({\em e.g}.~masses on springs in dash-pots) connected in {\em
parallel}. Each over-damped element has the characteristics of a Maxwellian fluid,
with a different time constant. Even if only a vanishingly small
fraction of the population of Maxwell models becomes infinitely viscous
({\em i.e.\ }their time constants $\overline{\tau}$ diverge), it will be
apparent in the response of the system, whose
stress is the {\em sum} of stresses of the population\footnote{The
connectivity of the damped oscillators, which is parallel for the GR model
(section \protect\ref{GR}), will be different in many real systems, becoming a
more elaborate network of series and parallel connections.}. Contrast this
with the GSER for a small probe,
Eq.\ \ref{wrong}. It claims to derive the moduli from a knowledge of only the
second moment, $\left<r^2\right>$, of the distribution of displacements. This is
(qualitatively) more like summing {\em compliances} of the
population\footnote{Of course, the {\em valid} application of the
GSER (to a large probe) {\em is} equivalent to {\em parallel} connection of
Maxwell models, since the surface of the large probe feels the
simultaneous influence of many degrees of freedom, which, within the
GR model, have additive stresses.}.

This discussion shows that the second moment alone,
$\left<r^2\right>$, does not contain
sufficient information about the distribution of displacements (or,
equivalently, about the full distributions $\Psi(\tau)$) to find the
rheological behaviour of the GR model. Specifically, this moment would be
unaffected by a vanishingly small subset of the population being stuck for
infinite time at $r=0$; yet this subset may dominate the rheology,
{\it e.g}.\ by causing $\left<\tau\right>_{\rm occ}$ to diverge.

\section{Hopping statistics in the GR model}
\label{analysis}

As we have mentioned, the GR model shows some intriguing dynamical
features. The viscosity diverges at a temperature ($T=2$) well above the
glass transition temperature ($T_{g}=1$ defined as the temperature below which
the system has no equilibrium steady state).
The diffusion constant $D$ vanishes at
$T_{g}$, and yet not all particles are static below this temperature. In this
section we gain a fuller picture of the model's dynamics,
by investigating its hopping statistics analytically and by
simulation\footnote{A similar analysis was carried out by Kehr et al.\
\protect\cite{Kehr78,Haus87} on a real lattice ({\it i.e}.\  with spatial
correlations). They modelled diffusion in crystals which exhibit no glass
transition, and therefore used a distribution of residence times narrower
than ours.}.

Let us begin by comparing the two distributions
$\Psi_{\rm hop}(\tau)$ and $\Psi_{\rm occ}(\tau)$, for the
residence times of particles in a trap. The first is taken over the prior
distribution (of {\it a priori} available states); the second over those
states which are actually occupied at any given time. As stated in section
\ref{test}, these two distributions are so different that selecting just the
first step on a random walk from the wrong distribution can have drastic
effects on $\left<r^2(t)\right>$ which persist up to rather late times.
Suppose that
by mistake the {\em initial} residence times were drawn from the
distribution $\Psi_{\rm hop}(\tau)$ (which is the correct choice for all
subsequent residence times), rather than from
$\Psi_{\rm occ}(\tau)$ (which is correct for the first one only
\cite{Kehr78,Haus87}).
The mean square displacement $\left<r^2(t)\right>_{\rm prior}$ over this
hypothetical ensemble has Laplace transform
\begin{equation}
\label{r2prior}
  \left<\widetilde{r^2}(s)\right>_{\rm prior} =
  \frac{\widetilde{\Psi}_{\rm hop}(s)\,d}{s\,(1-\widetilde{\Psi}_{\rm hop}(s))}
  + \frac{T\,d}{s}
\end{equation}
which, in contrast to Eq.\ \ref{r2s2}, does not depend only on the mean of
the distribution $\Psi_{\rm hop}(\tau)$. The resultant
$\left<r^2(t)\right>_{\rm prior}$ is a non-linear function of
time (tending to Eq.\ \ref{r2t} as $t\to\infty$ \cite{Bouchaud90}) which
increases monotonically, even at $T=T_{g}$ (where the
true diffusion constant vanishes).

The correct value of $\left<r^2(t)\right>$ given in Eq.\ \ref{r2t}
is not found as just
outlined, but instead by weighting each walk in the ensemble with the
residence time of the trap it is initially in \cite{Kehr78,Haus87}. This
factor is required because
the probability of a particle initially being found in a given trap is
proportional to the residence time of that trap. (Since we are interested in a
system which has already equilibrated, the particles have, in principle, had
time to sample the distribution of traps, and are therefore more likely to
be found in long-lived states than {\it a priori}.) It is somewhat
unintuitive (though necessary as a consequence of the system's steady state)
that this re-weighting of the ensemble should conspire to
cancel all non-linearity from the time dependence of mean square displacements
implicit in Eq.\ \ref{r2prior} and yield Eq.\ \ref{r2t} instead.

For the exponential prior
$\rho(E)=\exp(-E)$, it can be deduced straightforwardly
from Eqs.\ \ref{laplace} and \ref{averages} that, in Laplace time,
\begin{eqnarray*}
  \widetilde{\Psi}_{\rm hop}(s) &=&
    T\int_{0}^{\infty}\!\!\!\!du\, \frac{e^{-(T+1)u}}{s+e^{-u}} \\
  \widetilde{\Psi}_{\rm occ}(s) &=&
    (T-1)\int_{0}^{\infty}\!\!\!\!du\, \frac{e^{-Tu}}{s+e^{-u}}
\end{eqnarray*}
from which it follows, showing temperature dependence explicitly, that
\begin{equation}
\label{specialpsi}
  \Psi_{\rm occ}(\tau\,;\,T) = \Psi_{\rm hop}(\tau\,;\,T-1).
\end{equation}
Thus, at equilibrium, the residence time distribution of occupied states
coincides with the prior distribution of such times at a substantially lower
temperature.

\subsection{Numerical results}

To observe the process in action, we simulated the activated hopping
of $10^5$ particles, using the exponential prior distribution of trap
depths $\rho(E)=\exp(-E)$. For each particle, the time interval for
successive hops was drawn from $\Psi_{\rm hop}(\tau)$ (Eq.\
\ref{psioftau}) for all but the first hop (as in \cite{Kehr78}).
The algorithm used to simulate hopping in the system between times $0$ and
$t$ is as follows. Each of the $10^5$ particles is initially scheduled with a
hopping time drawn from $\Psi_{\rm occ}(\tau)$. If this is greater than $t$,
the particle is not visited again, and is recorded as performing zero
hops during the simulation. The remaining particles are successively
assigned further relaxation times\footnote{The stochastic variable
$\tau=-\xi^{-1/T}\ln\xi'$ is distributed according to $\Psi_{\rm hop}(\tau)$ if
$\xi$ and $\xi'$ are distributed uniformly on $(0,1)$.}
drawn from $\Psi_{\rm hop}(\tau)$ and their numbers of hops
incremented, until the next hop is
scheduled for after $t$. In this way the distribution $p_{t}(N)$ of the number
$N$ of relaxation events up to time $t$ was measured, as a function of
temperature.

\begin{figure}
  \epsfxsize=15cm
  \begin{center}
  \leavevmode\epsffile{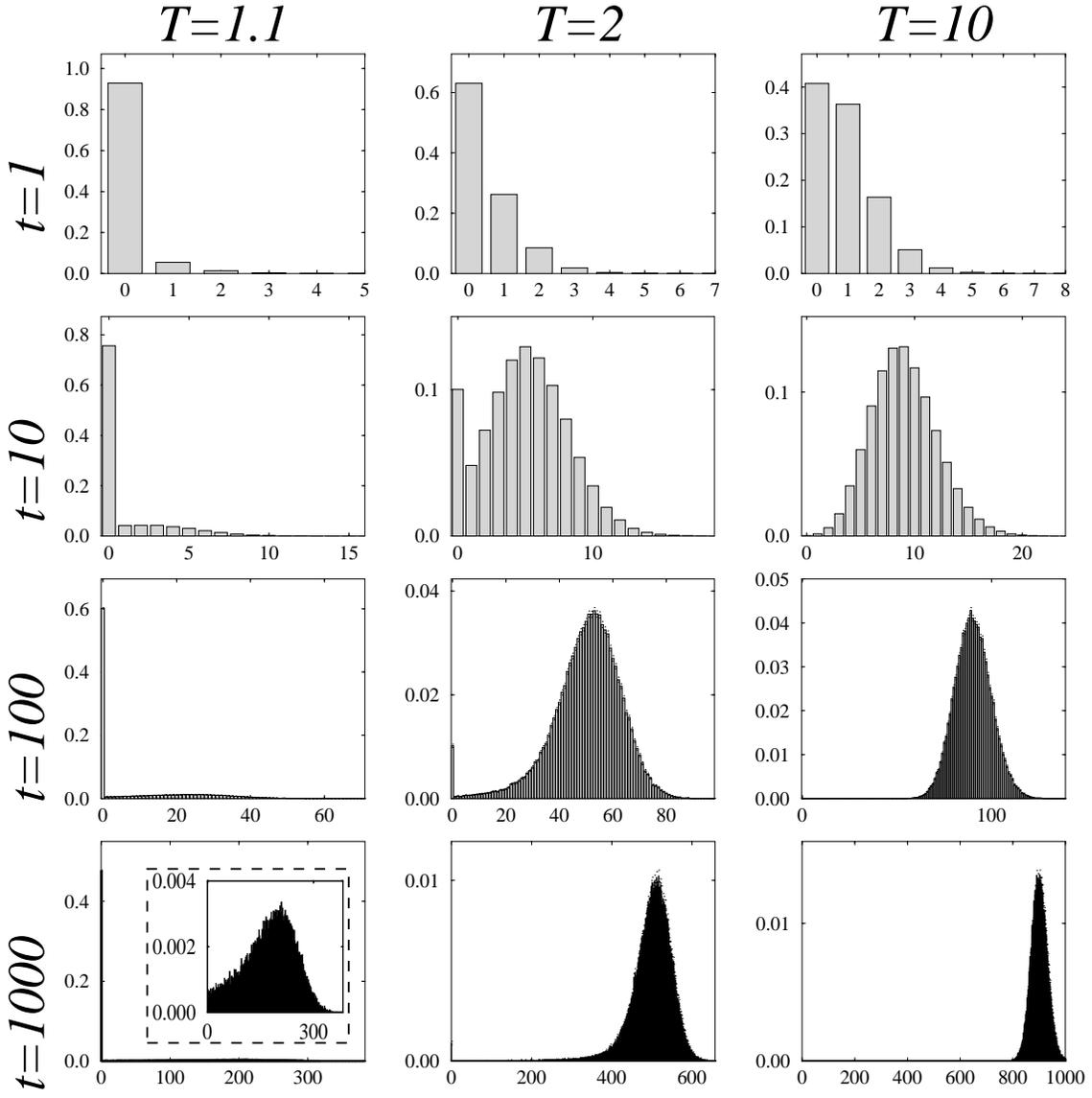}
  \caption{Normalized distributions of number of hops performed, $p_{t}(N)$,
           in simulations at various temperatures $T$ and times $t$. For
           $(T=1.1, t=1000)$, $p_{t}(0)=0.476 \pm 0.002$ and the rest of the
           distribution is magnified in the inset.}
  \label{p(N)}
  \end{center}
\end{figure}

The distribution of hops $p_{t}(N)$ gives a clearer picture of dynamics in the
system than the distribution of displacements, which is just its convolution
with a random walk for which $\left<r^2\right>=Nd$.
The results appear in Fig.\ \ref{p(N)}
for a range of temperatures and times. We see that, at high
temperature, all the particles are mobile and concentrated in a peak whose
position increases linearly with time. As the temperature is lowered
towards the glass transition ($T=1$), a second peak containing a
significant fraction of the population appears at $N=0$. Of course, as
this peak decays, it feeds the rest of the distribution.
Its lifetime increases as the transition is approached, yet the rest of the
distribution, for which it is the source, remains concentrated in a mobile
peak.

\subsection{Analysis of bimodal diffusion}

We consider next the time dependence of $p_{t}(0)$, which is the
probability that the time of the first hop exceeds $t$:
\begin{equation}
\label{p(0)}
  p_{t}(0) = \int_{t}^{\infty} \Psi_{\rm occ}(\tau)\,d\tau.
\end{equation}
Using Eq.\ \ref{specialpsi} and expressing $\Psi_{\rm hop}(\tau)$
in terms of an incomplete gamma function,
$\Psi_{\rm hop}(\tau) = \tau^{-(T+1)}\gamma(T+1,\tau)/T$
(from Eq.\ \ref{psioftau} with $\rho(E)=e^{-E}$), we have
\begin{equation}
\label{decay}
  p_{t}(0) = \frac{\gamma(T,t)}{t^{T-1}} + \frac{e^{-t}}{t}
  \; \tends{t}{\infty} \; (T+1)! \; t^{-(T-1)}.
\end{equation}
So the population of stuck particles decays by an increasingly slow power law
as the glass transition is approached. Of course, this decay is fastest at
early times, so this is when the maximum in the {\em mobile} peak is formed.
The width of the mobile peak grows with the usual square-root relation to its
mean so, at late times, the population is concentrated in two distinct and
narrow regions.

This confirms the result of the above simulation, that the diffusion process
in the GR model leads, near the glass transition, to a strongly bimodal
distribution of displacements. Having hopped once, a particle is more likely
to hop again in a given time, since relaxations in the mobile peak are drawn
from the (effectively higher-temperature --- Eq.\ \ref{specialpsi}) hop
distribution $\Psi_{\rm hop}(\tau)$. Nevertheless, the distribution of
residence times throughout the whole system (both peaks) is
$\Psi_{\rm occ}(\tau)$. Furthermore, the simulational data confirm that,
although the mean of the mobile peak is a non-trivial function of time and
temperature (cf.\ Eq.\
\ref{r2prior}), the two peaks together conspire to produce
\begin{displaymath}
  \overline{N} = \frac{t}{\left<\tau\right>_{\rm hop}} = \frac{T-1}{T}\:t
\end{displaymath}
in agreement with Eqs.\ \ref{r2t} and \ref{D}. In Ref.\ \cite{Kehr78}, such
linearity in time was shown to hold in the steady state of any system of
random walkers. Nevertheless, in our highly bimodal distribution, it appears
somewhat remarkable.

That the ``diffusion coefficient" $D$ should actually comprise an average over
two such distinct populations might call into question whether the dynamics of
the model can properly be called diffusion at all: the mean
square displacement may be linear in time, but is this diffusion in the
ordinary sense? To address the question, we calculate the {\em variance} in
$r^2(t)$, that is
$\left<r^4\right>-\left<r^2\right>^2$, and find, for $1<T<2$,
\begin{displaymath}
  \frac{\sqrt{\left<r^4\right>-\left<r^2\right>^2}}
	{\left<r^2\right>}\tends{t}{\infty}\sqrt{\frac{2}{d}}
	+ \frac{5\,\Gamma(T)}{(3-T)(2-T)\sqrt{2d}\;t^{T-1}}.
\end{displaymath}
The leading term is the standard value for simple diffusion. The way in which
this limit is approached {\em is} temperature dependent, but such sub-dominant
terms are unlikely to impinge on experimental measurements of the diffusion
constant in the ergodic ($T>1$) system. The deviation from a Gaussian
distribution of {\boldmath$r$} at finite times is often characterized by the
`non-Gaussian parameter' $A(t)$ defined \cite{Rahman62} as
\begin{displaymath}
  A(t)\equiv \frac{3}{5} \frac{\left< r^4\right>}{\left< r^2\right>^2} -1
\end{displaymath}
in 3 dimensions. Above the glass transition, this is expected
\cite{Odagaki94} to vanish in the infinite-time limit. For $1<T<2$, we find
\begin{displaymath}
  A(t) \tends{t}{\infty} \frac{2\Gamma(T)}{(3-T)(2-T)\; t^{T-1}}
\end{displaymath}
whereas $A \sim 1/t$ above $T=2$.
So the non-Gaussian parameter in the GR (and Bouchaud's) model does indeed
vanish, but does so increasingly slowly as the glass transition is
approached. The same qualitative results were observed by (amongst others)
Miyagawa et al.~in MD simulations \cite{Miyagawa91}.

\subsection{Timescales}

We conclude this section on unforced diffusion in the GR model by
summarizing the important time-scales that are present in both diffusion and
rheology. The diffusion is controlled by 
$\left<\tau\right>^{-1}_{\rm hop}$, equivalent
to the {\em mean rate}, $\left<\tau^{-1}\right>_{\rm occ}$, 
which vanishes linearly at
$T=T_{g}=1$. The {\em mean time} between yielding events in the system,
$\left<\tau\right>_{\rm occ}$ controls the GR model's viscosity, and diverges at
$T=2$, although the mean hopping rate is finite here. The {\em median} of
$\Psi_{\rm occ}(\tau)$ also remains finite below $T=2$. From Eq.\
\ref{p(0)}, it is the time at which half the population has relaxed (when
$p(0)=\mbox{$\frac{1}{2}$}$). From Eq.\ \ref{decay} this is
\begin{equation}
  t_{\mbox{$\frac{1}{2}$}}(T) \approx \left[ 2(T+1)! \right]^{\frac{1}{T-1}}
\end{equation}
which has a rather strong divergence at the glass transition ($T=1$).
Meanwhile, the {\em mode} of the distributions $\Psi_{\rm hop}(\tau)$ and
$\Psi_{\rm occ}(\tau)$, {\it i.e}.\ the most likely time between hops, is always
zero.

Given these subtle statistical properties of the self-diffusion process,
which arise from the power law distribution of relaxation times in the GR
model, it is no surprise that the GSER (which from Eq.\ \ref{D} contains only
$\left<\tau\right>_{\rm hop}$) fails for microscopic probes.

\section{Diffusion in the Presence of Shear}
\label{shear}

So far we have explored only the {\em linear} response functions of the GR
model, $G^*(\omega)$ and $\langle r^2(t)\rangle$, and their interrelation.
These describe any system for which the applied shear stress is either zero
or so small as not to disturb thermal equilibrium.  Let us now
turn our attention to applied shear rates which are sufficiently large to
influence self diffusion, and calculate the effect. Of course, particles in
Brownian motion are also convected by the system's affine shear motion, but
this effect (in which we include Taylor dispersion) can be subtracted
from the overall trajectory to reveal the perturbed diffusive
contribution\footnote{This applies, at least, in numerical studies, and was
done in a recent MD simulation
\protect\cite{Yamamoto98}. Taylor dispersion is the enhanced spreading in
the flow direction arising from particles diffusing from one streamline to
the next. When this and the trivial affine motion is subtracted, the remaining
Brownian motion had negligible anisotropy.}. (In simple shear, no such
subtraction is needed for the component of Brownian motion transverse to the
flow.) The rheology of the GR model \cite{Sollich97,Sollich98}
exhibits strong nonlinearities arising from shear-induced
hops. These occur as a particle is strained within its well, and moves up the
energy curve toward the yield value, making a thermal hop far more likely.
Clearly such shear-induced hops could have a radical effect on the diffusive
properties too. We assume that, although biased by the affine motion, each
such hop still entails a random displacement whose statistics is the same as
when flow is absent.

After subtraction of the affine motion, the diffusion constant is defined by
$\left<r^2(t)\right>\tends{t}{\infty}2Dtd$ in $d$ dimensions. For a random
walk of $N$ hops,
$\left<r^2\right>=Nd$ so, to calculate $D$, we just 
require the number of hops executed
by an average particle in a given long time. That is,
\begin{equation}
\label{diffusion}
  2D = \lim_{t\to\infty}\left< \frac{N}{t} \right> 
	= \frac{1}{\left<\tau\right>_{\rm hop}\!\!\!}
\end{equation}
since the time spent in each trap along the
walk\footnote{The hop-weighted average $\left<\tau\right>_{\rm hop}$,
which arises from the microscopic treatment, is completely equivalent to
the time-weighted average $\left<\tau^{-1}\right>^{-1}_{\rm occ}$ which arises from
the Fokker-Planck equation used in Refs.\ \protect\cite{Sollich97,Sollich98}. See
Eq.\ \protect\ref{averages}.} is drawn from
$\Psi_{\rm hop}(\tau)$, and each random walk becomes typical after a
sufficiently long time. We now evaluate Eq.\ \ref{diffusion} in the presence
of shear.

Given a time-dependent macroscopic shear $\gamma(t)$,
the yield rate for a trap which was entered at time $t_{0}$ was prescribed
in section \ref{GR} as
\begin{displaymath}
  \Gamma_{0} \exp\{-\beta(E-\mbox{$\frac{1}{2}$} k [\gamma(t)-\gamma(t_{0})]^2)\}
\end{displaymath}
since the trap has subsequently been sheared by an amount
$l=\gamma(t)-\gamma(t_{0})$. At time $t$, the probability of {\em not}
having hopped from the trap of depth $E$ entered at $t_{0}$ (which we call
$f_{E}(t,t_{0})$) is therefore
\begin{displaymath}
  f_{E}(t,t_{0}) = \exp \left\{
                - e^{-\beta E} \int_{t_{0}}^{t} \!\! dt'\,
                \exp\left( \mbox{$\frac{1}{2}$}\beta [\gamma(t')-\gamma(t_{0})]^2
                \right) \right\}
\end{displaymath}
where $\Gamma_{0}\equiv 1$ and $k\equiv 1$ as before.
Integrating over the prior distribution of trap
depths $\rho(E)$ gives the overall survival probability,
\begin{eqnarray}
\label{f}
  & & f(t,t_{0}) = \\
  & & \int_{0}^{\infty} \!\! dE \, \rho(E) \,
                \exp \left\{ - e^{-\beta E} \int_{t_{0}}^{t} \!\! dt'\,
                \exp\left( \mbox{$\frac{1}{2}$}\beta [\gamma(t')-\gamma(t_{0})]^2
                \right) \right\}.        \nonumber
\end{eqnarray}
The conditional probability density $\psi(t\,|\,t_{0})$ of
hopping next at time $t$, given that the present trap was entered at
$t_{0}$ is then
\begin{equation}
\label{conditional}
  \psi(t\,|\,t_{0}) = - \frac{d}{dt} f(t,t_{0}).
\end{equation}
Hence, the mean yield time, {\em given that the present trap was
entered at time $t_{0}$} is
\begin{eqnarray}
\label{timedependent}
  \left<\tau\right>_{t_{0}} &=& \int_{t_{0}}^{\infty} \!\! dt \,
        (t-t_{0})\, \psi(t\,|\,t_{0})           \nonumber \\
  &=& \int_{t_{0}}^{\infty} f(t,t_{0})\,dt.
\end{eqnarray}
We next apply Eqs.\ \ref{diffusion}, \ref{f} and \ref{timedependent} to
two particular forms of
$\gamma(t)$: steady shear, $\gamma(t) = \dot{\gamma}\,t$, and periodic
shear, $\gamma(t) = \gamma(t+2\pi/\omega)$.

\subsection{Steady Shear}
\label{steadyshear}

With a time-independent shear rate, the system reaches a steady state. (This
is true for all temperatures, including those below $T_g$: the presence of a
finite shear rate destroys weak ergodicity breaking \cite{Sollich97,Sollich98}.)
Therefore, the mean relaxation time becomes independent of the time of entry
into a trap. Thus Eqs.\ \ref{diffusion} and \ref{timedependent} are related by
\begin{displaymath}
   \left<\tau\right>_{\rm hop} = \left<\tau\right>_{t_{0}}
\end{displaymath}
for any $t_{0}$ we might choose. Thus we obtain
\begin{mathletters}
\begin{eqnarray}
  \label{steadyD}
    D^{-1} &=& 2\kappa^{-1}T \int_{0}^{\kappa^{-1}} \!\!\!\! du \,
                \rho(-T\ln\kappa u) \, \frac{J(u)}{u} \\
  \label{J}
    \mbox{with }\;J(u) &\equiv&  \int_{0}^{\infty} \!\! d\theta \,
				 e^{-u\,I(\theta)} \\
  \label{I}
    I(\theta) &\equiv& \int_{0}^{\theta} \!\! dv\,e^{v^2} \\
    \mbox{and }\;\;\;\;\;\kappa^2 &\equiv& \dot{\gamma}^2/2T
\end{eqnarray}
\end{mathletters}
from the substitutions $u^{-1}=\kappa\, e^{\beta E}$ and $\theta=\kappa\, t$.
Here, $\kappa$ is a re-scaled shear rate.

To proceed further we must approximate $J(u)$ (Eq.\ \ref{J}).
Let us define $\theta_{0}$ by $u I(\theta_{0})\equiv 1$.
Then for $\theta>\theta_{0}$, $u I(\theta)$ grows rapidly with $\theta$,
while for $0<\theta<\theta_{0}$ it is small. So we can write
\begin{displaymath}
  \int_{0}^{\infty} \!\! d\theta \, e^{-u\,I(\theta)}
  \approx \int_{0}^{\theta_{0}} \!\! d\theta = \theta_{0}(u).
\end{displaymath}
This approximation is equivalent to saying that shear does not greatly
modify the thermal activation rate while a particle is in the bottom of a
quadratic well, but yielding is immediate once the particle has been
sheared to the brink of its trap. The result is
\begin{equation}
\label{approxJ}
  J(u) \approx I^{-1}(1/u)
\end{equation}
where $I^{-1}$ is the inverse of $I$ in Eq.\ \ref{I}. We now approximate this
inverse function by
\begin{equation}
\label{approxI}
  I^{-1}(y) \approx I^{-1}_{\rm approx}(y) \equiv \left\{
  \begin{array}{lrl}
        y                               & \mbox{for } & 0\leq y<y_{0} \\
        1+\sqrt{\ln y} \hspace{12pt}    & \mbox{for } & y_{0}\leq y
  \end{array} \right.
\end{equation}
which is exact in the limits $y \ll 1$ and $y \gg 1$. The constant
$y_{0}\approx 1.747$ solves $y_{0}=1+\surd\ln y_{0}$ and is the
(second\footnote{Choosing the other solution, $y_{0}=1$, creates an
upturn in the gradient of $I^{-1}$, generating artefacts.})
point of intersection of the two approximate functions. In fact, it can
be shown that $I^{-1}_{\rm approx}(1/u)$ is always greater than $J(u)$
in Eq.\ \ref{steadyD} which it approximates. So, the
resulting approximation for $D$ is a strict {\em lower bound}.
General expressions for this are given in appendix
\ref{steadyapp}. Two regimes of the
scaled shear rate arise: $\kappa<y_{0}$ and $\kappa>y_{0}$.

\begin{figure}
  \epsfxsize=6.5cm
  \begin{center}
    \leavevmode\epsffile{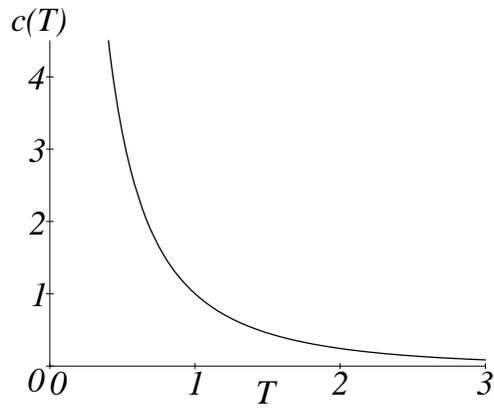} \vspace{1mm}
    \caption{The function appearing in Eqs.\ \protect\ref{bound} and
     \protect\ref{belowtrans}, $c(T) \equiv T^{-1} y_{0}^{1-T} -
     \frac{\sqrt{\pi}}{2} T^{-\frac{3}{2}}(T-1)\:\mbox{erfc}\sqrt{T\ln y_{0}}$,
     where erfc is the complementary error function and
     $y_{0}\approx1.747$.}
    \label{cfig}
  \end{center}
\end{figure}

Using the exponential prior $\rho(E)$ we find, in the limit of high shear
rate,
\begin{equation}
  D \tends{\kappa}{\infty} \frac{\kappa}{2\sqrt{\ln\kappa}}.
\end{equation}
So, for fast shear, the diffusion rate is almost proportional to the
shear rate. The imposed shear drags fluid elements rapidly through
trap configurations, with very little time for thermal activation. More
interesting behaviour occurs in the low-shear regime. With the exponential
prior, the approximate (lower bound) $D$ is (for $T<3$ --- see appendix
\ref{steadyapp})
\begin{equation}
  \label{bound}
    D \approx \frac{(T-1)}{2T} \left[ 1-\kappa^{T-1}c(T) \right]^{-1}
        \hspace{12pt} \mbox{ for } \kappa<y_{0}.
\end{equation}
The function $c(T)$ is given in Fig.\ \ref{cfig}.

Below the glass transition ($T<1$), the diffusion constant is zero in the
absence of shear, and particle motion is sub-diffusive, with $\left<r^2\right>$
growing as some non-integer power of time \cite{Monthus96}. However,
any finite steady shear rate ensures a steady rate of yielding
events and a non-zero diffusion constant. In this regime, from Eq.\
\ref{bound},
\begin{equation}
\label{belowtrans}
  D \tends{\kappa}{0} \frac{1-T}{2T\,c(T)}\: \kappa^{1-T}.
\end{equation}
Thus, $D$ is a finite, smooth function throughout all of the
($T$,$\dot{\gamma}$) plane, except for the line segment
$(\dot{\gamma}=0\, , \, 0 \leq T\leq 1)$, where it vanishes. For non-zero
shear rate, there is no singularity at $T=1$, where $D$ becomes
$-1/(2\ln\kappa)$ for low scaled shear rate $\kappa$.

\subsection{Periodic Shear}

A planar periodic shear strain, $\gamma(t)=\gamma(t+2\pi/\omega)$, is
experimentally more easily realizable than a constant shear rate. If we
define the phase $\phi$ to be $\omega t$ {\it modulo} $2\pi$, then we can
express the mean time $\left<\tau\right>_{t_{0}}$ spent in each trap on the walk
(defined in Eq.\ \ref{timedependent}) as a function,
$\overline{\tau}(\phi)$, only of the {\em phase} at which the current
trap was entered. This is done in appendix \ref{tau-and-p} by averaging
$\left<\tau\right>_{t_{0}}$ over all oscillatory cycles, for all times $t_{0}$
corresponding to phase $\phi$. If $p(\phi)$ is the
probability that a trap was entered at phase $\phi$ of the cycle, then the
mean residence time in Eq.\ \ref{diffusion} is
\begin{equation}
\label{tau-periodic}
  \left<\tau\right>_{\rm hop} = \int_{0}^{2\pi} \!\! d\phi \, p(\phi) \,
                        \overline{\tau}(\phi).
\end{equation}
Let $p(\phi\,|\,\phi_{0})$ be the {\em conditional} probability density
of a hop occurring at phase $\phi$, {\em given} that the previous hop was
at $\phi_{0}$. Then $p(\phi)$ is the (normalized) solution to the integral
equation
\begin{equation}
\label{Fredholm}
  p(\phi) = \int_{0}^{2\pi} \!\! p(\phi\,|\,\phi_{0}) \, p(\phi_{0}) \:
                d\phi_{0}.
\end{equation}
The quantities $p(\phi\,|\,\phi_{0})$ and $\overline{\tau}(\phi)$ have
simple relations to the distribution $\psi(t\,|\,t_{0})$
given in Eq.\ \ref{conditional}. In appendix \ref{tau-and-p} they are
given in terms of a single cycle of the imposed waveform.

Equation \ref{Fredholm} is a homogeneous Fredholm equation
of the second kind, for which the general solution is not known. However,
the problem greatly simplifies in the limit of small amplitude oscillations,
which we now treat. (Note that this perturbative treatment is only possible
for $T>1$, where an unperturbed steady state exists.) In the absence of shear
$p(\phi)$ and
$\overline{\tau}(\phi)$ must be independent of $\phi$. In that case, we write
$p(\phi)=1/2\pi$ and $\overline{\tau}(\phi)=\overline{\tau}_{0}$.
If the oscillatory shear
creates only small perturbations $\Delta p(\phi) = p(\phi)-1/2\pi$ and
$\Delta\overline{\tau}(\phi) = \overline{\tau}(\phi)-\overline{\tau}_{0}$,
then Eq.\ \ref{tau-periodic}, to first order in small
quantities, becomes
\begin{equation}
  \left<\tau\right>_{\rm hop} = \frac{1}{2\pi}
    \int_{0}^{2\pi} \overline{\tau}(\phi) \: d\phi.
\end{equation}
We have used the fact that $p$ is normalized, so that
$\int_{0}^{2\pi} \Delta p(\phi)\,d\phi = 0$. Hence we need only expand
Eq.\ \ref{tauexpression} to lowest order in strain amplitude and average it
over a cycle to find $\left<\tau\right>_{\rm hop}$ and hence the diffusion
constant. The result for an exponential prior $\rho(E)=\exp(-E)$ and
sinusoidal shear of amplitude $\gamma_{0}$ is
\begin{equation}
\label{sinusoidal}
  D^{-1} = \frac{2T}{T-1} - \gamma_{0}^2 \, \omega^{T-1}
        \int_{0}^{\omega^{-1}} \frac{u^{T-2}}{1+u^2}du
        + {\cal O}(\gamma_{0}^4)
\end{equation}
For $T<3$, the integral is convergent as $\omega\to 0$. In fact, for
{\em any} waveform, the low-frequency limit for an exponential prior is
\begin{equation}
  D^{-1} \tends{\omega}{0} \frac{2T}{T-1} - \gamma_{0}^2 \,
        \omega^{T-1} \, g(T) + {\cal O}(\gamma_{0}^4).
\end{equation}
where, in the sinusoidal case, $g(T)=(\pi/2)\,\mbox{cosec}\,[(T-1)\pi/2]$. In the
high-frequency limit, Eq.\ \ref{sinusoidal} yields instead
\begin{displaymath}
  D^{-1} \tends{\omega}{\infty} \frac{2T-\gamma_{0}^2}{T-1}
        + {\cal O}(\gamma_{0}^4).
\end{displaymath}

For larger amplitudes, we can use the results of section
\ref{steadyshear}. Given a sufficiently low frequency ($\omega \ll D$),
the shear rate at any given point on the cycle is almost constant. In that
case, the steady-shear expressions for $D^{-1}$ can be averaged over a
sinusoidal cycle, yielding (for $\omega \ll D$)
\begin{equation}
  D \approx \frac{T-1}{2T} \left[ 1 - (\omega\,\gamma_{0})^{T-1}
        h(T) \right]^{-1}; \hspace{5pt}
        \gamma_{0}^2 \omega^2 < 2T
\end{equation}
and, to within logarithmic factors,
\begin{equation}
  D \propto \omega\,\gamma_{0}\;\; ; \hspace{1.5in}
        \gamma_{0}^2 \omega^2 \gg 2T
\end{equation}
where $h(T)\equiv 2^{(T-1)/2} c(T)\, B(T/2,T/2)/\pi$, $c$ is given in
Fig.\ \ref{cfig} and $B$ is Euler's integral of the second kind.

We now summarize the effects, calculated above, of low-frequency periodic shear
on diffusion just above the glass transition. For the smallest
amplitudes $\gamma_{0}$, we expect the change in the diffusion constant to
be proportional to $\gamma_{0}^2 \, \omega^{T-1}$. At larger amplitude,
this should cross over to $(\gamma_{0}\,\omega)^{T-1}$, and at the largest
amplitudes, $D$ is almost proportional to $\omega\,\gamma_{0}$. We
interpret these three regimes as follows. The smallest oscillations serve
only to vibrate particles within the bottoms of their quadratic
potential wells, thus effectively making all wells just a little shallower.
Larger amplitude strains cause yielding of even the deepest wells in the
time taken for the affine shear to reach the yield point. However,
the part of the population in the shallowest wells are thermally activated
more quickly than this affine shear time-scale. The dividing line between
those two parts of the population depends on the shear rate. At the highest
shear rates, the global affine shear accounts for almost all yielding.

\section{Discussion and Conclusion}
\label{concl}
We have studied quantitatively, within a simple hopping model of glassy
dynamics, the relation between diffusion and rheological responses (in the
linear response regime) and (for the nonlinear regime) the coupling between
these two aspects of the dynamics.

In the presence of a broad distribution of relaxation times, as the model
possesses (and as is generic in hopping models of the glass transition) one can
expect strong violations of the generalized Stokes-Einstein relation (Eq.\
\ref{GSER}) when applied \cite{Mason95} to the self-diffusion of
representative particles in the medium (small probes). The relation holds only
for probe particles large enough that their surroundings are properly viewed
as a continuum \cite{Schnurr97}. Effective-medium theories, and also simple
forms of mode-coupling theory \cite{Mode} (in which a single mode of slow
relaxation dominates)
can give misleading predictions under these conditions. For glassy
systems, use of the GSER will under-predict the diffusivity of particles and/or
the rheological viscoelastic moduli. This is because the latter
are dominated by the most immobile and the former is dominated by the most
mobile particles.

In fact, the nature of diffusion in the model is quite
subtle (see section \ref{analysis}): close to (but
above) the glass transition one has an apparently bimodal behaviour.
Starting from any initial equilibrium state, a fraction of particles remain
stuck for a long time, but any particle that has hopped once remains mobile
thereafter. It is peculiar that, despite this, the diffusive behaviour
conspires to be relatively normal (at least for low order moments of the
displacement distribution; moments of fractional negative order would
presumably reveal a different story).

An additional peculiarity of diffusion, which our model does not
include, but which could also lead to enhanced diffusivity near the glass
temperature, has been observed in a recent molecular dynamics simulation
\cite{Donati98}. In Ref.\
\cite{Donati98} the effective dimensionality of the most mobile random walks
was observed to decrease with temperature. At low temperature, particles moved
along string-like clusters (also seen in Ref.\ \cite{Yamamoto98}), and
therefore covered greater distances than in an uncorrelated random walk in
three dimensions.

We have also applied our GR model to calculate the change in the
self-diffusion constant when a material is sheared. After
subtraction of affine motion (including Taylor dispersion) \cite{Yamamoto98},
or equivalently restricting attention to diffusion perpendicular to the shear
direction, one finds a strong effect of imposed flow on the mean jump rate
\cite{Sollich98} and hence on the diffusion constant. The effect is
particularly extreme below the glass transition, where an anomalous
(sub-diffusive) behaviour is converted to a finite diffusivity which has,
instead, a power law dependence on the steady shear rate. Such effects should
be accessible in scattering experiments (with wavevector almost
perpendicular to the flow direction) on labelled small probes. In oscillatory
shear,  an enhancement in $D$ was predicted in section \ref{shear} for
systems above (but near) the glass temperature; this could also be probed via
scattering.

The GR model as described in section \ref{GR} was originally developed in
Ref.\ \cite{Sollich97} to reproduce the generic rheology of a class of ``soft
glassy materials". (We comment further below on its relevance to conventional
glasses.) This class  was argued to include, for example, foams, emulsions,
pastes and slurries. Experimentally, their linear viscoelastic behaviour is
often characterized by a nearly constant ratio of the elastic and loss moduli
$G'(\omega)$,
$G''(\omega)$ ($G''/G'$ is usually about 0.1) with a frequency dependence that
is either a weak power law (clay slurries, paints, microgels) or
negligible (tomato paste, dense emulsions, dense multi-layer
vesicles) \cite{Mackley94}. This behaviour persists down to the
lowest experimentally accessible frequencies. Sometimes a regime is
seen at small $\omega$ where $G'$ is constant and $G''$ is decreasing
(which can be interpreted in terms of the model's behaviour below $T_g$
\cite{Sollich97,Sollich98,Fielding}).

As mentioned in sections \ref{introd} and \ref{GR}, there are two differences
between the GR model used in this paper and its soft counterpart in
\cite{Sollich97}. The first is that the {\em soft} GR model refers not to
particles, as we have done, but
to mesoscopic material ``elements", large enough for a local elastic strain
variable to be defined but small enough to have strong heterogeneity in local
yield energies. (For the case of a foam, say, an ``element" could correspond
to a domain of several bubbles, and ``yield" to a local topological change.) It
is not necessarily clear what is meant by self-diffusion of such elements, so
in the present paper we have retained a particle picture, although this is more
natural for conventional glass-forming liquids than for soft glassy materials.

The second difference is that in the {\em soft} GR model, $T$ is replaced by an
effective ``noise temperature" $x$. It was argued in \cite{Sollich97} that the
resemblance to thermal activation is formal: the ``activated" yield processes
are viewed as arising primarily by coupling to structural rearrangements
elsewhere in the system. Indeed, the elastic energies associated with local
rearrangements in foams and the like are many orders of magnitude in excess of
$k_{B}T$, so any interpretation of $x$ as a true temperature is somewhat
unconvincing for these materials (in contrast to conventional glasses).
Although this interpretation remains problematical, as discussed in
\cite{Sollich97,Sollich98}, with it the soft GR model is able to reproduce many of
the rheological properties of soft glassy materials.

The results we have obtained for the GR model, concerning breakdown of the
GSER and the nature of shear-induced diffusion, equally apply (with $T\to x$)
to the {\em soft} GR model. The breakdown of GSER, arising from the fact that
diffusion and rheology probe different aspects of the relaxation spectrum, is
equally natural in this case; the other contribution to its breakdown
discussed in \ref{breakdown}, arising from the temperature dependence of the
local attempt frequency, even more so (since $x$ is anyway not a true
temperature). The results of section \ref{shear} for shear-induced diffusion
could also be quite interesting for soft glassy materials, in which it is easy
to apply shear
strains large enough to strongly perturb the intrinsic relaxation times. The
length scales in these materials can be probed via light scattering;  index
matching is often possible so that true tracer diffusion (of a small subset of
unmatched droplets or particles) can be measured \cite{index}. For oscillatory
flows, an important innovation is the echo technique \cite{echo} in which the
positions of a given scatter at identical points in the shear cycle are
compared.

Results on dense emulsions \cite{echo} suggest, in fact, that not only
are shear-induced reorganizations easily detectable, but that these have
strong temporal and spatial correlations --- regions that reorganize at a given
point in the shear cycle will do so again in the next one. Such correlations
are not
included in the (soft) GR model(s) although they might be added in principle
\cite{Falk}. (It would require escape from a shallow trap to be preferentially
into another shallow trap, though we do not advocate appending such ad-hoc
correlations to this simplified model.) It would be very
interesting to know whether the same applies in conventional glasses;
preliminary work on colloidal suspensions (which are traditionally
thought of in these terms) suggests not \cite{Haw}.

Finally, we return to the phenomenology of the GR model in relation to
conventional glass-forming liquids. As explained in Section \ref{GR}B, the GR
model in its basic form predicts a viscosity divergence at $T=2$, while the
diffusion constant vanishes only at $T=1$, where the system has a glass
transition to a non-ergodic state. This existence of a temperature range
with finite diffusivity but infinite viscosity appears to be at odds with
the experimentally observed behaviour of conventional glasses. A common
divergence of viscosity and inverse diffusion constant can however be
incorporated into the GR model through a cutoff $E_{\rm max}$ on the
distribution of yield energies $\rho(E)$; this modification of the model has
in fact already been discussed in Ref.~\cite{Sollich98}.
It yields a viscosity which initially follows the original power-law
divergence as $T=2$ is approached, but then crosses over to
$\eta \sim \exp(E_{\rm max}/T)$ as $T$ is lowered further. Similarly, the
predicted $1/D$ would first seem to diverge as $T=1$ is approached from
above, but actually remain finite there and eventually approach infinity at
the same temperature as the viscosity ($T=0$). The shear moduli obey the
power laws (Eq.~\ref{powerlaws}) down to a cutoff frequency
$\omega_{\rm min}=\exp (-E_{\rm max}/T)$, but then cross over to low-frequency
Maxwell behaviour ($G' \sim \omega^2$, $G'' \sim \omega$).

The above simple temperature dependences apply if, as we did throughout,
we assume that the (prior) density of yield energies is temperature
independent. This is of course an approximation; Odagaki, for example,
suggested that the width of $\rho(E)$ may in fact scale as the inverse of the
amount of free volume $v(T)$ in the system \cite{Odagaki95}. In all
temperature dependences, $T\,v(T)$ then replaces $T$. As a consequence, if the
free volume decreases smoothly to zero at a finite temperature $T_{\rm VF}$,
a Vogel-Fulcher-like divergence of $\eta$ and $1/D$ at $T_{\rm VF}$ (rather
than the above Arrhenius behaviour) would be predicted by the GR model with
energy cutoff. One would then be inclined to locate the glass transition at
that point; if we revert to temperature independent $\rho(E)$, this
corresponds to $T=0$ (rather than $T=1$, which is the appropriate choice in
the absence of an energy cutoff). Results in the range $0<T<1$ for the GR
model with cutoff may therefore actually apply to supercooled liquids
{\em above} the glass transition. These include a dynamic modulus
$G^* (\omega)$ which becomes {\em more} Maxwellian in the low-frequency range
as $T$ is lowered (for $T<1$, one has $G' \sim \mbox{const}$.\ and
$G'' \sim \omega^{T-1}$ above the cutoff frequency $\omega_{\rm min}$
\cite{Sollich98}, and hence Maxwell behaviour with relaxation time
$1/\omega_{\rm min}$ for $T \to 0$), in an intriguing correspondence with data
taken by Menon et al.~\cite{Menon94}.

\begin{small}
This work was funded by EPSRC Grant No.\ GR/K56025 (RMLE) and a Royal Society
Dorothy Hodgkin fellowship (PS). We wish to thank P. N. Pusey, F. Lequeux, P.
H\'{e}braud and J.-P. Bouchaud for helpful discussions.
\end{small}

\appendix
\section{General Expressions for the Diffusion Constant Under Steady Shear}
\label{steadyapp}

Substituting Eqs.\ \ref{approxJ} and \ref{approxI} into \ref{steadyD} yields
lower bounds on $D$ (upper bounds on $D^{-1}$) in two regimes:
\begin{eqnarray}
  D^{-1} &<& 2\kappa^{-1}T \int_{0}^{\kappa^{-1}} \!\!\!\!\!\! du \,
        \frac{\rho(-T\ln\kappa u)}{u} \, \left(1+\sqrt{\ln u^{-1}}\right)
        \nonumber \\
  & & \hspace{1.75in} \mbox{ for } \kappa > y_{0}  \\
  D^{-1} &<& 2\kappa^{-1}T \int_{0}^{y_{0}^{-1}} \!\!\!\! du \,
        \frac{\rho(-T\ln\kappa u)}{u} \, \left(1+\sqrt{\ln u^{-1}}\right)
        \nonumber \\
        &+& 2\kappa^{-1}T \int_{y_{0}^{-1}}^{\kappa^{-1}} \!\!\!\! du \,
        \frac{\rho(-T\ln\kappa u)}{u^2}
        \hspace{12pt} \mbox{ for } \kappa < y_{0}.
\end{eqnarray}
In each case, $D$ is approximately equal to the lower bound for $T<3$.

\section{Evaluation of the Phase-Dependent Relaxation Time and
Conditional Probability under Periodic Strain}
\label{tau-and-p}

The probability of leaving a trap at time $t$ given that it was
entered at $t_{0}$ (at phase $\phi_{0}=\omega t_{0} \,\mbox{mod}\, 2\pi$
of the oscillatory
cycle) is $\psi(t\,|\,t_{0})$. Hence the mean time spent in the trap, given
that it was entered at phase $\phi_{0}$, is
\begin{displaymath}
  \overline{\tau}(\phi_{0}) = \int_{\phi_{0}/\omega}^{\infty} \!\!
        (t-\phi_{0}/\omega) \;\, \psi(t\,|\,\phi_{0}/\omega) \; dt
\end{displaymath}
which, from Eq.\ \ref{conditional}, gives
\begin{displaymath}
  \overline{\tau}(\phi_{0}) = \int_{\phi_{0}/\omega}^{\infty} \!\!
        f(t,\phi_{0}/\omega) \, dt
\end{displaymath}
with $f$ defined in Eq.\ \ref{f}. For compact notation, we define
$a(\omega t)\equiv\gamma(t)/\sqrt{2T}$ and $\tau_{E}(\phi_{0})$
according to
\begin{displaymath}
  \overline{\tau}(\phi_{0}) \equiv \int_{0}^{\infty} \!\! \rho(E) \,
        \tau_{E}(\phi_{0}) \, dE.
\end{displaymath}
Thus we find
\begin{eqnarray}
\label{workings}
  \omega\left.\tau_{E}(\phi_{0})\right|_{E=-T\ln\omega u} &=& \\
  & & \hspace{-15mm}
  \int_{\phi_{0}}^{\infty} \!\!\!\! d\phi \, \exp \left\{ \!-u
        \int_{\phi_{0}}^{\phi} \!\!\! d\phi' \,
        \exp[a(\phi')-a(\phi_{0})]^2 \right\} \nonumber
\end{eqnarray}
Since $\exp[a(\phi')-a(\phi_{0})]^2$ is a periodic function of $\phi$,
the integrand in Eq.\ \ref{workings} increases by a constant factor with
each cycle $\phi\to\phi+2\pi$. Thus the R.H.S. is a geometric series of
integrals over a single cycle,
\begin{eqnarray*}
  & & \left\{ \sum_{n=0}^{\infty} \left[ \exp \left( -u
        \int_{0}^{2\pi} \!\!\! d\phi' \, \exp[a(\phi')-a(\phi_{0})]^2
        \right) \right]^n \right\} \, \times \\
  & & \int_{\phi_{0}}^{\phi_{0}+2\pi} \!\!\!\!\! d\phi \, \exp \left\{ -u
        \int_{\phi_{0}}^{\phi} \!\! d\phi' \,
        \exp[a(\phi')-a(\phi_{0})]^2 \right\}.
\end{eqnarray*}
Finally, $\overline{\tau}(\phi_{0})$ is expressed in terms of a single
cycle as
\begin{eqnarray}
\label{tauexpression}
  \overline{\tau}(\phi_{0}) &=& \frac{T}{\omega} \int_{0}^{\omega^{-1}}
      \!\! du \, \frac{\rho(-T\ln\omega u)}{u} \int_{0}^{2\pi}
      \!\!\!\!\!\! d\phi \, \times \\
  & & \frac{ \exp\left\{-u\int_{\phi_{0}}
        ^{\phi+2\pi\Theta(\phi_{0}-\phi)} \!\! d\phi' \,
             \exp [ a(\phi')-a(\phi_{0}) ]^2 \right\} }
           { 1 - \exp\left\{ -u\int_{0}^{2\pi} \!\! d\phi' \,
             \exp [ a(\phi')-a(\phi_{0}) ]^2 \right\} }         \nonumber
\end{eqnarray}
where $\Theta(\phi_{0}-\phi)$ is the Heaviside step function.

Similarly,
\begin{displaymath}
  p(\phi\,|\,\phi_{0}) = \int_{\phi_{0}/\omega}^{\infty} \!\!\!
        \delta([\omega t \,\mbox{mod}\, 2\pi]-\phi) \;\,
        \psi(t\,|\,\phi_{0}/\omega) \; dt
\end{displaymath}
(with $\delta(x)$ the Dirac delta function) is also a geometric series,
from which it follows
\begin{eqnarray}
\label{pexpression}
  p(\phi\,|\,\phi_{0}) &=& T \int_{0}^{\omega^{-1}} \!\! du \,
        \rho(-T\ln\omega u) \, \frac{d}{d\phi} \, \times \\
  & & \frac{ \exp\left\{-u\int_{\phi_{0}}
        ^{\phi+2\pi\Theta(\phi_{0}-\phi)} \!\! d\phi' \,
             \exp [ a(\phi')-a(\phi_{0}) ]^2 \right\} }
           { 1 - \exp\left\{ -u\int_{0}^{2\pi} \!\! d\phi' \,
             \exp [ a(\phi')-a(\phi_{0}) ]^2 \right\} }.        \nonumber
\end{eqnarray}

\end{document}